\documentclass{llncs}
\usepackage{algorithm}
\usepackage{algpseudocode}
\usepackage{listings}
\usepackage{fancyhdr}
\usepackage{indentfirst}
\usepackage{graphicx}
\usepackage{amsmath}
\usepackage{amssymb}
\usepackage{graphicx}
\usepackage{epsfig}
\usepackage{amsmath}
\usepackage{amssymb}
\usepackage[all]{xy}
\usepackage{graphicx}
\usepackage{epsfig}
\usepackage{booktabs}
\usepackage{stmaryrd}
\usepackage{url}
\usepackage{longtable}
\usepackage[figuresright]{rotating}
\usepackage{multirow}
\usepackage{pifont}
\usepackage{longtable}
\usepackage{arydshln}
\usepackage[all]{xy}
\algblock{ParFor}{EndParFor}
\algnewcommand\algorithmicparfor{\textbf{parfor}}
\algnewcommand\algorithmicpardo{\textbf{do}}
\algnewcommand\algorithmicendparfor{\textbf{end\ parfor}}
\algrenewtext{ParFor}[1]{\algorithmicparfor\ #1\ \algorithmicpardo}
\algrenewtext{EndParFor}{\algorithmicendparfor}

\begin{document}
\title{A Parallel Memetic Algorithm to Solve the Vehicle Routing Problem with Time Windows}

\author{Jakub Nalepa\inst{1} \and Zbigniew J. Czech\inst{1,2}}

\authorrunning{J. Nalepa and Z.J. Czech}

\institute{Silesian University of Technology, Gliwice, Poland\\
\email{\{jakub.nalepa, zbigniew.czech\}@polsl.pl}
\and University of Silesia, Sosnowiec, Poland}
\maketitle

\begin{abstract}
This paper presents a parallel memetic algorithm for solving the vehicle routing problem with time windows (VRPTW). The VRPTW is a well-known NP-hard discrete optimization problem with two objectives. The main objective is to minimize the number of vehicles serving customers scattered on the map, and the second one is to minimize the total distance traveled by the vehicles. Here, the fleet size is minimized in the first phase of the proposed method using the parallel heuristic algorithm (PHA), and the traveled distance is minimized in the second phase by the parallel memetic algorithm (PMA). In both parallel algorithms, the parallel components co-operate periodically in order to exchange the best solutions found so far. An extensive experimental study performed on the Gehring and Homberger's benchmark proves the high convergence capabilities and robustness of both PHA and PMA. Also, we present the speedup analysis of the PMA.
\end{abstract}


\section{Introduction and Related Work}
The vehicle routing problem with time windows (VRPTW) consists in finding a schedule for a fleet of homogeneous vehicles serving a set of geographically scattered customers. The capacities of the vehicles cannot be exceeded and the customers must be visited within their well-defined time windows. We consider the VRPTW as a hierarchical optimization problem. The primary objective is to minimize the total fleet size and the second one is to minimize the total distance traveled by the vehicles. This approach makes it possible to develop and optimize the algorithms for both phases independently.

The practical applications of the VRPTW include the bus route planning, post and parcels delivering, food delivering, cash delivering to banks and ATM terminals and many more. Thus, a number of algorithms including exact and heuristic methods were introduced for the VRPTW over the years. Due to the large complexity of the VRPTW and its wide practical applicability the heuristic and metaheuristic methods capable of producing high-quality feasible solutions in reasonable time are of main importance here.

Exact algorithms incorporating, among others, dynamic programming, branch-and-bound algorithms and greedy approaches, were proposed by Bard et al.~\cite{Bard2002}, Irnich and Villeneuve~\cite{Irnich2006}, Jepsen et al.~\cite{Jepsen2006}, Kellehauge et al.~\cite{Kallehauge2006} and Chabrier~\cite{Chabrier2006}. Noteworthy, only 9 instances out of 300 belonging to the Gehring and Homberger's benchmark~\cite{HombergerBenchmark} have been solved to optimality~\cite{Nagata2010}. An extensive review of the exact methods can be found in Kallehauge~\cite{Kallehauge2008Review}.

Heuristic algorithms can be divided into two classes, namely the improvement and the construction techniques. In the construction heuristic algorithms the customers are iteratively inserted into a partial feasible solution without violating the time windows and capacity constraints. Several construction heuristics were proposed by Solomon~\cite{Solomon1987}, Potvin and Rousseau~\cite{Potvin1993} and recently by Pang~\cite{Pang2011}. On the other hand, the improvement heuristics modify an initial solution and explore the search space by performing local search moves in order to decrease the fleet size and the total traveled distance. The examples of such heuristics can be found in Thompson et al.~\cite{Thompson1993}, Russell~\cite{Russell1995} and Potvin and Rousseau~\cite{Potvin1995}.

Metaheuristic algorithms incorporate mechanisms to explore the search space and to exploit its most promising regions. They allow infeasible intermediate solutions and the solutions deteriorating during the search process in order to escape the local minima. A number of sequential and parallel algorithms were introduced during the recent years. The simulated annealing was successfully applied by Zhong and Pan~\cite{Zhong2007}, Debudaj-Grabysz and Czech~\cite{Debudaj2005}, and recently by Li~\cite{Li2013}. The tabu searches were proposed by Cordeau et al.~\cite{Cordeau2001} and Sin et al.~\cite{Sin2002}. The ant colony approaches can be found in Xuan et al.~\cite{Xuan2006} and Qi and Sun~\cite{Qi2008}. A survey on the metaheuristic algorithms can be found in Br\"{a}ysy and Gendreau~\cite{Braysy2005Survey}.

Evolutionary algorithms have attracted the scientific attention to solve the VRPTW due to their high search capabilities. Genetic algorithms were applied by Cheng and Wang~\cite{Cheng2009}, Kamkar et al.~\cite{Kamkar2010} and Ursani et al.~\cite{Ursani2011}. The evolution strategies were proposed by Gehring and Homberger~\cite{Gehring2002}, Mester et al.~\cite{Mester2005} and Kanoh et al.~\cite{Kanoh2010}. The sequential and parallel memetic algorithms (MAs), combining the evolutionary algorithms for more distant search with the local optimization and refinements algorithms for local exploitation, were described by Berger and Barkaoui~\cite{Berger2004}, Labadi et al.~\cite{Labadi2008} and Nagata and Br\"{a}ysy~\cite{Nagata2010}. The MA based on the edge-assembly crossover operator (EAX) has been proposed by Nagata and Br\"{a}ysy~\cite{Nagata2010}.

This paper is organized as follows. Section~\ref{sec:formulation} formulates the VRPTW. The sequential and the parallel heuristic algorithms for minimizing the number of routes are discussed in Section~\ref{sec:first_phase_algorithms}. The memetic algorithm for the total distance minimization is described in Section 4. Section 5 describes the experimental results. Section 6 concludes the paper.

\section{Problem Formulation} \label{sec:formulation}

Let $G=(V,E)$ be a directed graph with a set $V$ of $N+1$ vertices representing the customers and the depot, together with a set of edges $E=\{(v_i,v_{i+1})|v_i,v_{i+1}\in V, v_i \neq v_{i+1}\}$ representing the connections between the travel points. Each vehicle starts and finishes at the depot $v_0$. Travel costs are given as $c_{i,j}$, where $i \neq j$, $i,j \in \{0,1,...,N\}$. Non-negative customer demands $d_i$, $i \in \{0,1,...,N\}$, where $d_0=0$ and the time windows $[e_i, l_i]$, $i \in \{0,1,...,N\}$ are defined. The customers have their service times $s_i$, $i \in \{1,2,...,N\}$. A fleet of $K$ vehicles with a constant capacity $Q$ is given. The route is defined as a set of customers served by a single vehicle $(v_0,v_1,...,v_{n+1})$, where $v_0 = v_{n+1}$ is the depot.

A solution $\sigma$ is feasible if (i) the total amount of goods delivered to the customers within each route does not exceed the vehicle capacity $Q$, (ii) the service of a customer $v_i$ is started before the time window $[e_i, l_i]$ elapses, (iii) every customer $v_i$ is served in exactly one route, and (iv) every vehicle leaves and returns to the depot within its time window $[e_0, l_0]$.

The primary objective of the VRPTW is to minimize the fleet size $K$ ($K\leq K_{min}$, and $K_{min}=\left\lceil D/Q\right\rceil$, where $D=\sum_{i=1}^N{d_i}$). The secondary objective is to minimize the total distance $T=\sum_{i=1}^{K} T_i$, where $K$ is the number of routes and $T_i$ is the distance of the $i$-\textit{th} route.

\section{Minimizing the Number of Routes by the Parallel Heuristic Algorithm} \label{sec:first_phase_algorithms}

In this section we discuss in detail the heuristic algorithm used for minimizing the number of routes in the VRPTW. First, its sequential version is outlined along with the proposed enhancements. Then, the parallel heuristic algorithm (PHA) is presented.

\subsection{Sequential Algorithm Outline} \label{sec:seq_first_phase}

The route minimization heuristic algorithm reduces the number of routes in a feasible solution $\sigma$ by one at a time. The process of removing routes continues until the execution time reaches a defined maximum or the number of vehicles reaches its lower bound $K_{min}$~\cite{Nagata2009}. The initial solution of the VRPTW consists of $N$ routes, which means that each customer is served by a separate vehicle. A randomly selected route (Fig.~\ref{fig:sequential_first_phase}, line~\ref{alg:select_random_route}) is excluded from the current solution. The customers are inserted into the ejection pool (EP) (line~\ref{alg:put_ejected_into_EP}). The penalty counters $p[i]$, are set to 1 (line~\ref{alg:init_penalty}). The counters indicate the reinsertion difficulties of the corresponding customers. The customers are taken from the EP applying the LIFO strategy.

\begin{figure}[t]
\centering
\begin{algorithmic}[1]
%\State Create an initial solution $\sigma_{init}$;\label{alg:init_first_stage}
%\State $\sigma \gets \sigma_{init}$;
%\State \textit{finished} $\gets$ \textbf{false};
%\While{\textbf{not} finished}\label{alg:start_main_while}
\State Select and remove a random route $s$ from $\sigma$;\label{alg:select_random_route}
\State Put a random permutation of ejected customers of the removed route into EP;\label{alg:put_ejected_into_EP}
\State Initialize penalty counters $p[i] \gets 1, i = 1, 2,\dots, N$;\label{alg:init_penalty}
\While{(EP$\neq \emptyset$ \textbf{and} \textbf{not} \textit{finished})}\label{alg:start_inner_while}
\State Select and remove customer $v_{ins}$ from EP using LIFO strategy;\label{alg:get_cust_from_EP}
\If{$N^{f}_{ins}(v_{ins}, \sigma) \neq \emptyset$}\label{alg:check_Nf}
\State $\sigma$ $\gets$ $\sigma'$ chosen randomly from $N^{f}_{ins}(v_{ins}, \sigma)$;\label{alg:get_sigma_prime}
\Else \State $\sigma$ $\gets$ Squeeze($v_{ins}$, $\sigma$);\label{alg:squeeze}
\EndIf
\If{$v_{ins}$ $\notin$ $\sigma$}\label{alg:check_if_cust_in_sigma}
\State $p[v_{ins}] \gets p[v_{ins}]+1$;\label{alg:increase_penalty}
\State $\sigma$ $\gets$ $\sigma^{\prime}\in N^{\it ej\/}(v_{ins}, \sigma)$
       with $\min(\sum_{i=1}^{k} p[v^{(i)}_{out}])$;\label{alg:select_cust_to_eject}
\State Insert ejected customers $v^{(1)}_{out}$, $v^{(2)}_{out}$, \dots,
              $v^{(k)}_{out}$ into EP;\label{alg:insert_ejected_into_EP}
\State $\sigma$ $\gets$ Perturb($\sigma$);\label{alg:perturb}
\EndIf
\State \textit{finished} $\gets$ VerifyStopCondition();\label{alg:verify_stop}
\EndWhile
\If{EP$\neq \emptyset$}\label{alg:check_if_partial}
\State Restore $\sigma$ to the initial solution;\label{alg:restore_init_sigma}
\EndIf
%\EndWhile
\State \Return $\sigma$;\label{alg:return_sigma}
\end{algorithmic}
  \caption
  {
    A sequential heuristic algorithm to minimize the number of routes in the VRPTW (\emph{RemoveRoute}).
  }
  \label{fig:sequential_first_phase}
\end{figure}

All feasible insertion positions are determined (line~\ref{alg:check_Nf}). Introducing forward and backward time window penalty slacks allowed for constant-time verification of the time window penalties~\cite{Nagata2009} (clearly, the change of the vehicle loads may be computed in constant time). If the set of feasible insertion positions is not empty, then a random insertion is performed. The solution $\sigma'$ with the customer $v_{ins}$ becomes a new feasible (possibly partial) solution $\sigma$.

If the set $N_{ins}^f(v_{ins},\sigma)$ is empty, then it is impossible to insert the customer $v_{ins}$ without violating the constraints. The \emph{Squeeze} procedure allows for creating temporarily infeasible solutions (line~\ref{alg:squeeze}). The infeasible solution with the smallest value of the penalty function $F_p(\sigma)$ is chosen~\cite{Nagata2010}, and the attempts of restoring the feasibility are undertaken. If the squeezing fails, then the penalty counter of the customer is increased (line~\ref{alg:increase_penalty}).

The last approach of reinserting $v_{ins}$ allows for ejecting other customers from the solution. A sum of the penalty counters $P_{sum}$ is minimized to eject the customers that will be relatively easy to reinsert later. The customers that were inserted during the last $l_{max}$ iterations are not considered for ejections~\cite{Blocho2010}. Noteworthy, the increasing number of the ejected customers $k$, $k \in\{1,2,\dots,k_{max}\}$, is considered. If at least one feasible ejection is found for a given $k$, then the other tests are skipped. If there are more ejections with the same $P_{sum}$, then one is chosen randomly~\cite{Blocho2010}. A number of constant-time local moves are performed in \emph{Perturb} procedure in order to diversify the search (line~\ref{alg:perturb}). The algorithm finishes if the EP is empty, the execution time exceeds the specified limit or the size of the EP is unacceptably large (line~\ref{alg:verify_stop}). The other breaking conditions are described in the next section.

\subsection{Suggested Modifications} \label{sec:modifications}

The additional stopping condition of the algorithm introduced in~\cite{Blocho2010} addresses the maximal number of iterations $i_{max}$. According to that, it should break even though the size of the EP is small and $i>i_{max}$, where $i$ is the number of current iteration. It is worth noting that a large number of iterations are usually performed if the number of routes is close to the optimum. Thus, it may be proficient to allow for additional loop executions if the EP is small and the probability of reinserting the customers is still high. The additional parameter $\xi$ indicates the maximal number of customers allowed to reside in the EP for which the loop will continue despite of exceeding $i_{max}$.

However, the size of the EP can stay constant during the execution for a long time. Here, the maximal number of iterations in the steady state, $\psi$, is introduced. It should vary with the maximal number of allowed iterations, thus $\psi$ is a fraction of $i_{max}$. If the EP size does not change during the $\psi$ iterations, then the probability of a feasible customer reinsertion drops and the loop breaks.

In the \emph{Squeeze} procedure the feasibility restoring attempts are undertaken. The local search moves that may be calculated in constant time were used during the construction of $\mathcal{N}_r (\sigma)$. If there are no feasible moves, then the linear-time moves are considered. This approach is appropriate for smaller instances, since calculating the moves may become time-consuming for the larger number of customers. Additionally, it may be worth limiting the number of moves to test, e.g., until $n$ edge-exchanges are found.

A route to be removed is chosen randomly. The routes may be divided into two classes -- the first containing the routes with the number of customers greater or equal to the average, and the second class with the other routes. Intuitively, it should be easier to reinsert customers from the smaller route to a partial solution, thus choosing a random route from the first class gives a higher probability of feasible reinsertions. However, it is proficient to get rid of larger routes earlier and choose a random route from the second class, when the solution size, i.e., the number of routes, is still far from the optimum.

If the squeezing fails, then the other ejections are tested. After a successful reinsertion of $v_{ins}$ and ejections of other customers the solution is perturbed. In many cases perturbing is not necessary for efficient reinsertions. The perturbation may be omitted if the percentage of successful customer insertions without additional ejections is significant, e.g., $80\%$, for a given number of subsequent iterations. However, the algorithm may fail due to exceeding the maximal number of trials. In this case, the perturbations of the partial solutions should be allowed. The number of moves in the \emph{Perturb} procedure may depend on the difficulty of reinserting the customers, i.e., it should increase with the decrease of the number of routes. The initial number of moves is multiplied by a constant factor, e.g., 2, every defined number of iterations until it reaches the maximal value.

\subsection{Parallel Algorithm Outline} \label{sec:operators}

The $p$ available processors may be used either to achieve a higher accuracy of a solution in a given time, or to speed up the computations. In the first case, the goal is to obtain a solution that is closer to the global optimum. The main goal of the parallel heuristic algorithm (PHA) is to improve the accuracy of final solutions. The algorithm consists of $p$ components denoted as $P_0,P_1,\dots,P_{p-1}$\footnote[1]{In the OpenMP implementation we consider each thread as a parallel component.}. The initial solution is considered as the starting solution for each parallel component (Fig.~\ref{fig:parallel_first_phase}, line~\ref{alg:init_sigma_in_parallel}). Then, each component calls \emph{RemoveRoute} in parallel with others $\delta$ times. The components co-operate to exchange the best solutions found up-to-date. The solutions are assessed according to their costs. The solutions with the smaller number of routes $K$ are preferred. If the number of routes is equal, then the solution with the shorter total travel distance is considered better. The parallel algorithm guides the search towards the optimal solutions with respect to the number of vehicles and the travel distance.

\begin{figure}[t]

\centering
\begin{algorithmic}[1]
\ParFor{$P_i \gets$ $P_0$ \textbf{to} $P_{p-1}$}\label{alg:begin_outer_parfor}
\State $\sigma_i \gets \sigma_{init}$;\label{alg:init_sigma_in_parallel}
\EndParFor
\While{\textbf{not} \textit{finished}}\label{alg:start_while_pha}\label{alg:begin_while_in_pha}
\ParFor{$P_i \gets$ $P_0$ \textbf{to} $P_{p-1}$}\label{alg:begin_main_parfor_in_pha}
\State Call \emph{RemoveRoute} $\delta$ times;\Comment{See Fig.~\ref{fig:sequential_first_phase}}\label{alg:remove_route}
\EndParFor
\State Co-operate according to the co-operation scheme;\label{alg:cooperate_in_pha}
\State \textit{finished} $\gets$ VerifyStopCondition();
\EndWhile
\end{algorithmic}
  \caption
  {
    A parallel heuristic algorithm (PHA) to minimize the number of routes in the VRPTW.
  }
  \label{fig:parallel_first_phase}
\end{figure}

\subsection{Co-operation of Parallel Components}

\begin{figure}
\begin{displaymath}
\sigma_{init} \to \begin{cases}
    \xymatrix@C=1em@R=1em{ \sigma_{0}^{t_0} \ar[r] & \sigma_{0}^{t_1} \ar[r] \ar[d] & \sigma_{0}^{t_2} \ar[d] \ar[r] & \dots \ar[r] & \sigma_{0}^{t_{n-1}} \ar[r] \ar[d] & \sigma_{0}^{t_n} \ar[d]\\
\sigma_{1}^{t_0} \ar[r]& \sigma_{1}^{t_1} \ar[r] \ar[d] & \sigma_{1}^{t_2} \ar[d] \ar[r] & \dots \ar[r] & \sigma_{1}^{t_{n-1}} \ar[r] \ar[d] & \sigma_{1}^{t_n} \ar[d]\\
 &  &  &  &  &  \\
\dots & \dots & \dots & \dots & \dots & \dots \\
\sigma_{p-2}^{t_0} \ar[r] & \sigma_{p-2}^{t_1} \ar[r] \ar[d] & \sigma_{p-2}^{t_2} \ar[d] \ar[r] & \dots \ar[r] & \sigma_{p-2}^{t_{n-1}} \ar[r] \ar[d] & \sigma_{p-2}^{t_n} \ar[d]\\
\sigma_{p-1}^{t_0} \ar[r]& \sigma_{p-1}^{t_1} \ar[r] & \sigma_{p-1}^{t_2} \ar[r] & \dots \ar[r] & \sigma_{p-1}^{t_{n-1}} \ar[r] & \sigma_{p-1}^{t_n}}
\end{cases}
\end{displaymath}
\caption{Co-operation scheme}
\label{cooperationscheme}
\end{figure}
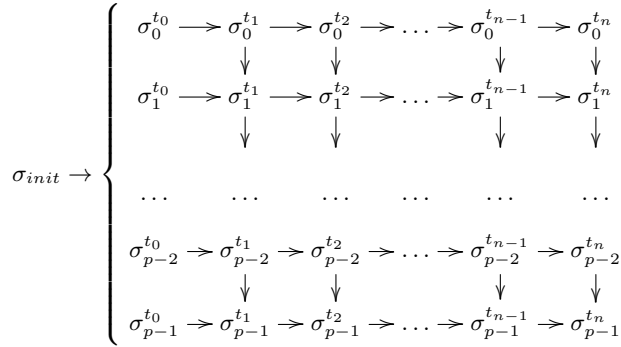

The co-operation of threads starts from thread $P_0$. Thread $P_1$ receives the solution $\sigma_0$ from $P_0$, and compares the costs of solution $\sigma_1$ with the received one. The solution with the smaller cost replaces the current solution of $P_1$. Consequently, thread $P_{p-1}$ compares the solution $\sigma_{p-1}$ with $\sigma_{p-2}$ received from $P_{p-2}$. Finally, the best solution is held by thread $P_{p-1}$. If the best solution is found by $P_0$, then all the components get $\sigma_0$. The co-operation scheme is presented in Fig.~\ref{cooperationscheme}.

The second variant of the co-operation scheme is cyclic and includes the communication between threads $P_{p-1}$ and $P_0$. The better solution will be sent to thread $P_0$. The solution $\sigma_0$ should be updated only if the number of routes is larger than the number of routes in $P_{p-1}$. Keeping the original solution $\sigma_0$ with the same number of vehicles by thread $P_{0}$ may prevent from having the same solution held by all threads. Introducing the probability of replacing a worse solution by the solution received from the neighbor may further decrease the likelihood of having only one solution in the team. The probability of choosing the better solution should be large, but not equal to 1. The number of steps that are executed in parallel before the co-operation must be determined sensitively. If the co-operation is too frequent, e.g. for large problem instances, the total parallel overhead becomes more significant and the execution time increases rapidly. However, if the components co-operate rarely for small instances, the gain from parallelism is hardly noticeable.

The following co-operation frequencies are introduced:
\begin{enumerate}
\item \textbf{Constant} -- $\delta$ is constant.
\item \textbf{Rare} -- $\delta$ is defined as a ratio of the problem size and a value of rare co-operation factor. Then, after a number of co-operation phases $\delta$ is divided by this factor until it reaches a defined lower limit.
\item \textbf{Frequent} -- the scheme is similar to the rare co-operation, but the frequent co-operation factor is larger than the rare co-operation factor.
\item \textbf{Adaptive} -- $\delta$ is defined as a ratio of the problem size and a value of adaptive co-operation factor. The $\delta$ is divided by the ratio of the last average time and the previous average time (if $t_{avg}^{prev} \neq 0$, otherwise $\delta$ is divided by the adaptive factor) of \emph{RemoveRoute} executions.
\end{enumerate}

An extensive study on the co-operation schemes can be found in~\cite{Nalepa2012TAAI}.

\section{Minimizing the Total Travel Distance by the Parallel Memetic Algorithm} \label{sec:exp}

Here, we discuss both sequential and parallel versions of the memetic algorithm for minimizing the total distance traveled by the vehicles. The main goal of the presented parallel algorithm is to reduce the execution time without decreasing the quality of final solutions.

\subsection{Sequential Algorithm Outline} \label{sec:exp}

The initial population of size $N$ containing the feasible solutions with $K$ routes is found by the parallel heuristic algorithm discussed in Section~\ref{sec:first_phase_algorithms}. If the maximal time of generating the initial population is exceeded, then the solutions already found are copied and perturbed until the population size reaches $N$.

Each individual is chosen once as the parent $p_A$ and $p_B$ in a random order to generate the child solutions using the EAX operator (Fig.~\ref{fig:sequential_second_phase}, line~\ref{alg:determine_N_pairs}) using the AB-selection scheme. It has been shown that the proper pre- and post-selection schemes have strong impact on the convergence capabilities of a memetic algorithm~\cite{Kawulok2012SPR}.

$N_{ch}$ defines the number of children generated for each pair $p_A$ and $p_B$. The feasibility of a child is restored by the \emph{Repair} function if necessary (line~\ref{alg:Repair}) using the concept of local moves utilized while squeezing an infeasible solution in the route minimization heuristics. If the solution is feasible, then a number of moves are performed to improve its quality, i.e., to decrease the total travel distance (line~\ref{alg:LocalSearch}). The moves are limited to the customers belonging to the routes modified by the EAX operator and the repairing procedure~\cite{Nagata2010}. The total distance of a new solution is compared with the total distance of the best child found up-to-date (line~\ref{alg:p_c_versus_best}). If a new solution is of higher quality, then the best child is updated.

After generation of child solutions for $N$ pairs of parents the population is updated, i.e., the best individuals form a new population according to the post-selection scheme. It is easy to see that the best child obtained for $p_A$ and $p_B$ replaces the first parent, not the worst individual in the population. Removal of $p_A$ is motivated by the fact that the better individual replaces the worse with the similar characteristics to ensure the population diversity. The additional termination condition addresses the steady state, i.e., the situation in which for a large number of subsequent generations the quality of the best individual is not improved. The algorithm finishes if the number of generations in the steady state is larger than the defined maximum, maximal number of generations $G$ is reached, or the maximal execution time is exceeded. The best individual from the population is finally returned (line~\ref{alg:return_best_individual}).

\begin{figure}[t]
\centering
\begin{algorithmic}[1]
\State Generate an initial population of $N$ feasible solutions;\label{alg:init_pop}
\While{\textbf{not} $finished$}\label{alg:start_main_loop}
\State Determine $N$ reproduction pairs ($p_A$, $p_B$);\label{alg:determine_N_pairs} \Comment{Pre-selection}
\For{each pair ($p_A$, $p_B$)}\label{alg:start_foreach_loop}
\For{i $\gets$ 1 \textbf{to} $N_{ch}$}\label{alg:foreach_child}
\State $p_{c}\gets$ EAX($p_A$, $p_B$);\label{alg:EAX}
\State $p_{c}\gets$ Repair($p_{c}$);\label{alg:Repair}
\State $p_{c}'\gets$ LocalSearch($p_{c}$);\label{alg:LocalSearch}
\If {$\eta(p_{c}')>\eta(p_{c}^b)$}\label{alg:p_c_versus_best}
\State $p_{c}^b\gets$ $p_{c}'$;\label{alg:update_best}
\EndIf\label{alg:p_c_versus_best_end}
\EndFor
\EndFor
\State Form the next population;\label{alg:form_new_population} \Comment{Post-selection}
\State $finished\gets$ VerifyTerminationCondition();\label{alg:verify_termination}
\EndWhile\label{alg:end_main_loop}
\\\Return best individual in the last population;\label{alg:return_best_individual}

\end{algorithmic}
  \caption
  {
    The memetic algorithm for the total distance minimization.
  }
  \label{fig:sequential_second_phase}
\end{figure}

\subsection{Parallel Algorithm Outline} \label{sec:exp}

The parallel memetic algorithm (PMA) consists of $p$ parallel components denoted as $P_0,P_1,\dots,P_{p-1}$. The main part of the memetic algorithm, i.e., generating the child solutions, is the most computationally intensive. The iterations of the loop may be executed in parallel, since $N_{ch}$ children are generated for the parents $p_A$ and $p_B$ independently. The best child solution is stored as $\sigma_{r(i)}^{best}$. Each individual in the population serves once as $p_A$ and $p_B$ during the combination stage, therefore different $\sigma_{r(i)}^{best}$ solutions are updated in every iteration. The $N$ iterations are distributed between $p$ threads, where $N \gg p$. The number of individuals in the population is usually large to avoid the similarities between the individuals. Once the loop finishes, the best child solutions are found and the current generation is updated. The cost, i.e., the total travel distance, of each individual in the current solution is compared with the cost of the best child. If the cost of the child $\sigma_i^{best}$ is smaller, then the child becomes a new individual in the population and replaces the solution $\sigma_i$. Clearly, the $N$ solutions are compared independently, therefore the iterations may be executed in parallel. Processing of the next generation of solutions starts with initializing of the set of the best child solutions. Similarly, the loop iterations are independent and may be executed in parallel.

\section{Experimental Results} \label{sec:exp}

The algorithms were implemented in C++ using the OpenMP interface and were tested on Gehring and Homberger's problem instances. The code was compiled using Intel C++ Compiler 10.1.015 with \texttt{-fast} and \texttt{-openmp} flags. Calculations were carried out at a single node of \emph{Galera} supercomputer at the Academic Computer Center in Gda\'nsk~\cite{galera}. The computations were performed on the nodes with 16 GB RAM (2 GB/core) equipped with Intel Xeon Quad Core (2.33 GHz) processors with 12 MB of level 3 cache. The parameters used during the experiments are given in Tab.~\ref{tab:params} and Tab.~\ref{tab:params2}. The percentage of the nearest customers $\mu$ is limited for neighborhood calculations to decrease the execution time~\cite{Nagata2009}. The number of additional customers allowed to reside in the EP has been proposed in~\cite{Blocho2010}. The minimal number of local moves used during the solution perturbation should allow transforming a current solution to the neighboring, but still not too similar one. If the additional ejections are necessary for a successful customer insertion, then the number of moves is multiplied by $I_P^f$ to increase the probability of getting the new configurations. The maximal number of moves (for both stages) prevents from a rapid increase of the execution time. The maximal number of iterations in the steady state corresponds to a decent fraction of the maximal number of allowed algorithm iterations.

\begin{table}[h!]
  \caption{
    Parameters of the route minimization heuristic algorithm.
  }\label{tab:params}

\renewcommand{\tabcolsep}{1mm}
\centering
\begin{tabular}{l l l l l}
\hline

Parameter &  & Description &  & Value\\
\hline

$\mu$   &    &  Percentage of the nearest customers in the \\
			      &    &neighborhood &  &  $0.6$ \\
$k_{max}$   &    &  Maximal number of customers to be ejected &  &  $3$ \\
$l_{max}$   &    &  Number of iterations without ejecting a customer \\
			      &    & after it is inserted &  &  $5$ \\
$\xi$   &    &  Additional customers allowed to reside in the EP &  &  $7$ \\
$i_{max}$   &    &  Maximal number of iterations in the first phase \\
			      &    &algorithm &  &  $1000$ \\
$\psi$   &    &  Maximal number of iterations in the steady state &  &  $i_{max}/5$ \\
$I_P^{m}$   &    &  Minimal number of feasible moves\\
			      &    & while perturbing &  &  $80$ \\
$I_P^{M}$   &    &  Maximal number of feasible moves \\
			      &    &while perturbing &  &  $400$ \\
$I_P^{f}$   &    &  Update factor for the number of moves \\
			      &    &while perturbing &  &  $2$ \\
$I_F$   &    &  Frequency of updating the number of moves \\
			      &    &(in iterations) &  &  $50$ \\
$\tau_R$   &    &  Maximal time for reinsertions in \emph{RemoveRoute} \\
			      &    &(in seconds) &  &  $50$ \\
$\tau$   &    &  Maximal execution time (in seconds) &  &  $1200$ \\

\hline

\end{tabular}
\end{table}

\begin{table}[h!]
  \caption{
    Parameters of the distance minimization memetic algorithm.
  }\label{tab:params2}

\renewcommand{\tabcolsep}{1mm}
\centering
\begin{tabular}{l l l l l}
\hline

Parameter &  & Description &  & Value\\
\hline

$N_{ch}$   &    &  Number of child solutions generated for each pair of \\
			      &    &parents &  &  $20$ \\
$I_C$   &    &  Maximal number of moves improving the child solution &  &  $100$ \\
$I_P$   &    & Number of moves used during copying and perturbing  &  &  $50$ \\
$\tau$   &    &  Maximal execution time (in minutes) &  &  $N_sN/400$ \\
$G$   &    &  Maximal number of generations without improvement  & &  $50$ \\

\hline

\end{tabular}
\end{table}

The settings of the co-operation are given in Tab.~\ref{tab:params3}. The EAX strategy~\cite{Nagata2010} is chosen randomly. If a significant number of consecutive generations, e.g., 50, does not result in improving the best individual in a population, then the probability of further improvements drops rapidly. A formula for the maximal execution time calculation of the memetic algorithm has been proposed in~\cite{Nagata2010}.

\begin{table}[h!]
  \caption{
    Co-operation frequency settings; CM -- co-operation mode, CF -- co-operation factor, UF -- update factor, Ufr -- update frequency, Mfr -- minimal frequency.
  }\label{tab:params3}

\renewcommand{\tabcolsep}{1mm}
\centering
\begin{tabular}{l l l l l l l l l l l l}
\hline

Cust. no. &  & CM &  & CF & & UF & & Ufr & & Mfr\\
\hline

200 & & Frequent & & 10 & & 2 & & 4 & & 1 \\
400 & & Frequent & & 10 & & 2 & & 4 & & 1 \\
600 & & Adaptive & & 10 & & -- & & 1 & & 1 \\
800 & & Rare & & 5 & & 2 & & 3 & & 1 \\
1000 & & Rare & & 5 & & 2 & & 3 & & 1 \\

\hline

\end{tabular}
\end{table}

\begin{figure}[t!]
\centering

\renewcommand{\tabcolsep}{0cm}
\newcommand{\myfigwidth}{0.4}
\newcommand{\raiseshift}{0.1mm}
\newcommand{\mytextsize}{\scriptsize}

\begin{tabular}{ll}

a) & b) \\
\includegraphics[height=\myfigwidth\columnwidth]{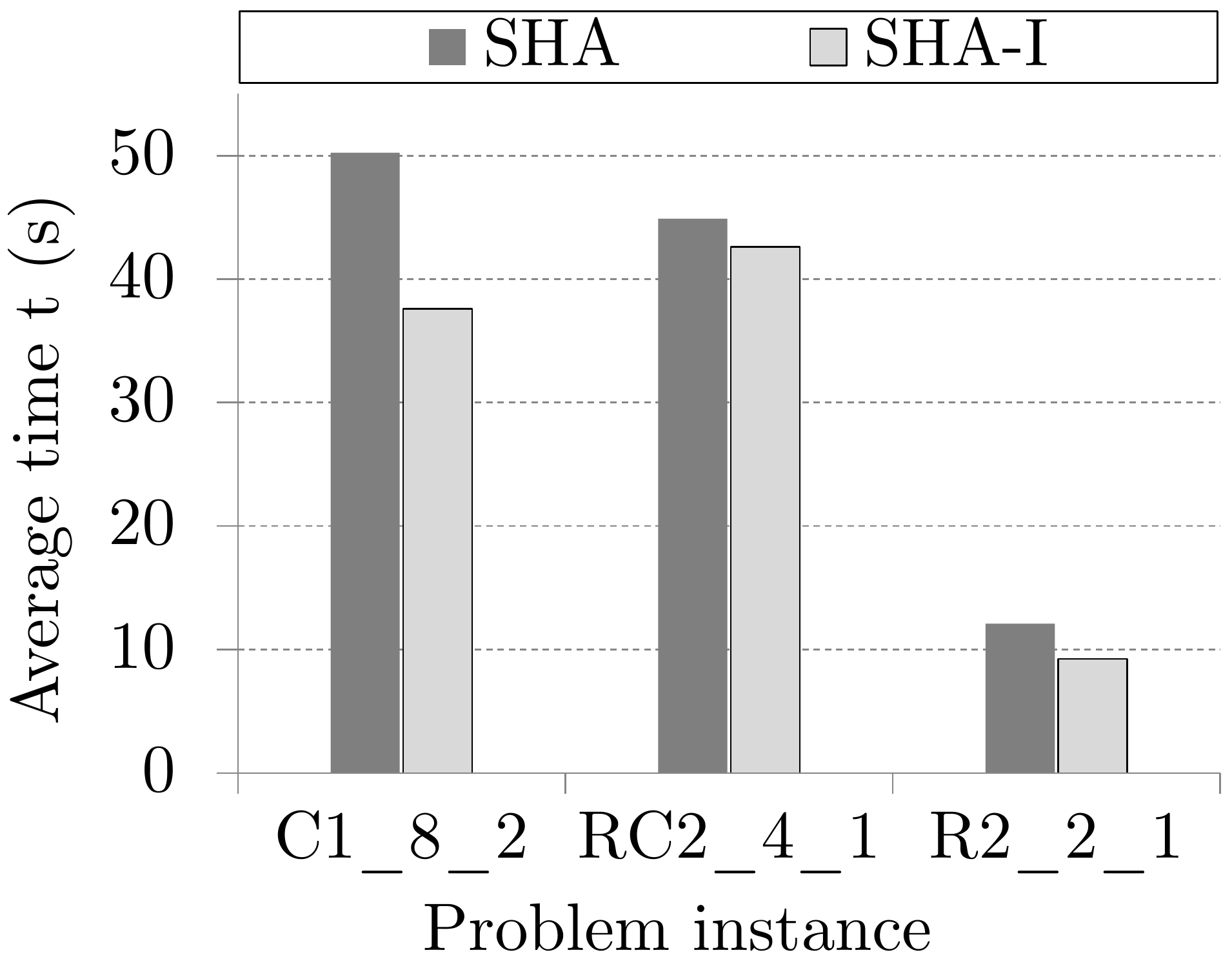} \hspace{0.01\columnwidth}&
\includegraphics[height=\myfigwidth\columnwidth]{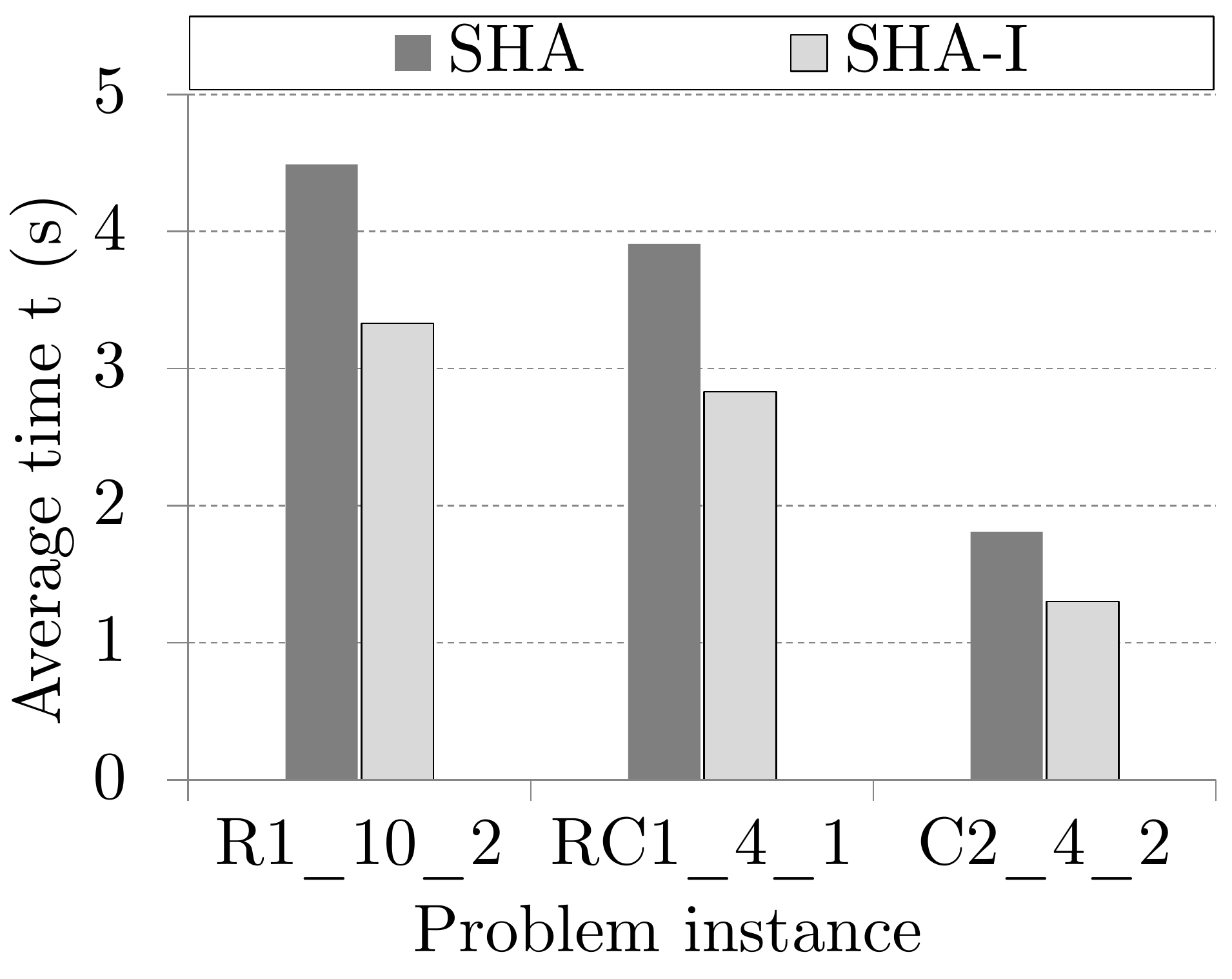}
\\

\end{tabular}

  \caption
  {
    Average execution time $t$ (in seconds) of a sequential heuristic algorithm (SHA) and the improved sequential heuristic algorithm (SHA--I) for minimizing the number of routes for more (a) and less (b) time-consuming tests for 100 experiments.
  }
  \label{fig:avg_exec_times}
\end{figure}

A number of possible modifications and improvements have been suggested in Section~\ref{sec:modifications}. The exemplary average execution times of the sequential route minimization heuristic algorithm are given in Fig.~\ref{fig:avg_exec_times}. If the algorithm gets stuck in the local minima of the search space (e.g., for RC2\_4\_1), then the decreased initial number of local search moves results in the increase of the total number of iterations necessary to leave the local minimum. However, it is not always necessary to explore the vast solution space for large instances (e.g., for R1\_10\_2, C1\_8\_2) and a relatively small number of moves during the perturbation is enough to get satisfactory results. The average execution time has been decreased for a number of instances that were relatively easy to solve (Fig.~\ref{fig:avg_exec_times}b) and for time-consuming ones (a). However, the modifications are less suitable for the problems with solution spaces containing a large number of local minima.

The cumulative numbers of vehicles (CVNs), i.e. the number of vehicles serving all instances, are presented in Tab.~\ref{tab:cvns} for sequential and parallel algorithms (SHA--I and PHA, respectively). The number of vehicles was decreased for 16 instances, whereas the world's best results were obtained in 6 cases using the parallel heuristics. Therefore, in 271 out of 300 (90\%) cases the benchmarking tests were solved to the current optimum with respect to the number of vehicles using the parallel algorithm. The parallel memetic algorithm significantly improved the current world's best result for the problem instance C1\_8\_2\footnote[2]{The best-known solutions of the GH tests are published at: \url{http://www.sintef.no/Projectweb/TOP/VRPTW/Homberger-benchmark/}; reference date: September 13, 2011.}.

\begin{table}[h!]
  \caption{
    The percentage of the world's best CVNs obtained using SHA--I and PHA.
  }\label{tab:cvns}

\renewcommand{\tabcolsep}{1mm}
\centering
\begin{tabular}{l l l l l l l l l l l l}
\hline

Class &  & SHA--I & & PHA\\
\hline

C1 & & 82\% & & 84\% \\
C2 & & 70\% & & 78\% \\
R1 & & 94\% & & 94\% \\
R2 & & 100\% & & 100\% \\
RC1 & & 100\% & & 100\% \\
RC2 & & 84\% & & 86\% \\
\hdashline
Total & & 88\% & & 90\%\\
\hline

\end{tabular}
\end{table}

\begin{figure}[t!]
\centering

\renewcommand{\tabcolsep}{0cm}
\newcommand{\myfigwidth}{0.38}
\newcommand{\raiseshift}{0.1mm}
\newcommand{\mytextsize}{\scriptsize}

\begin{tabular}{ll}

a) & b) \\
\includegraphics[height=\myfigwidth\columnwidth]{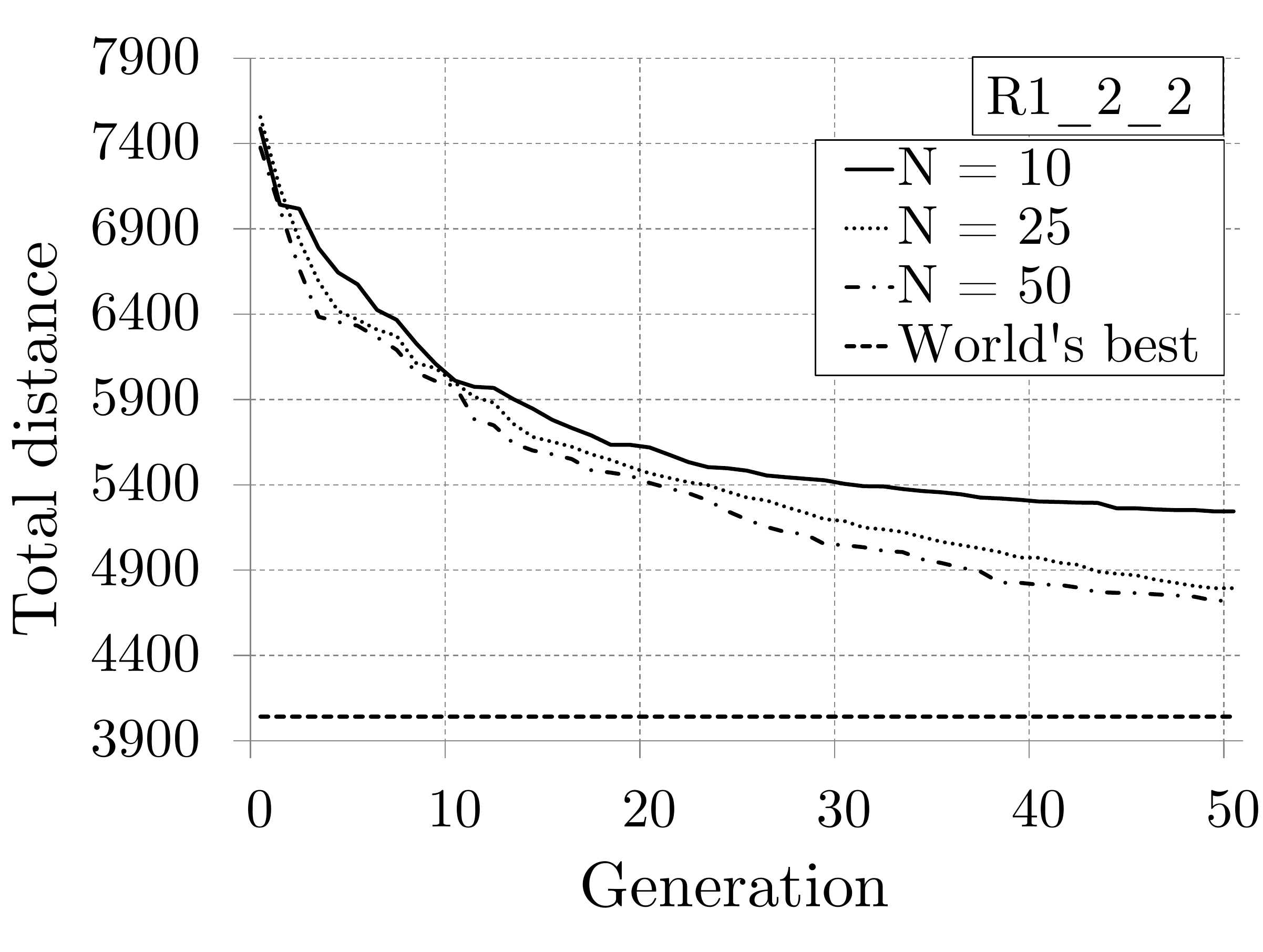} \hspace{0.01\columnwidth}&
\includegraphics[height=\myfigwidth\columnwidth]{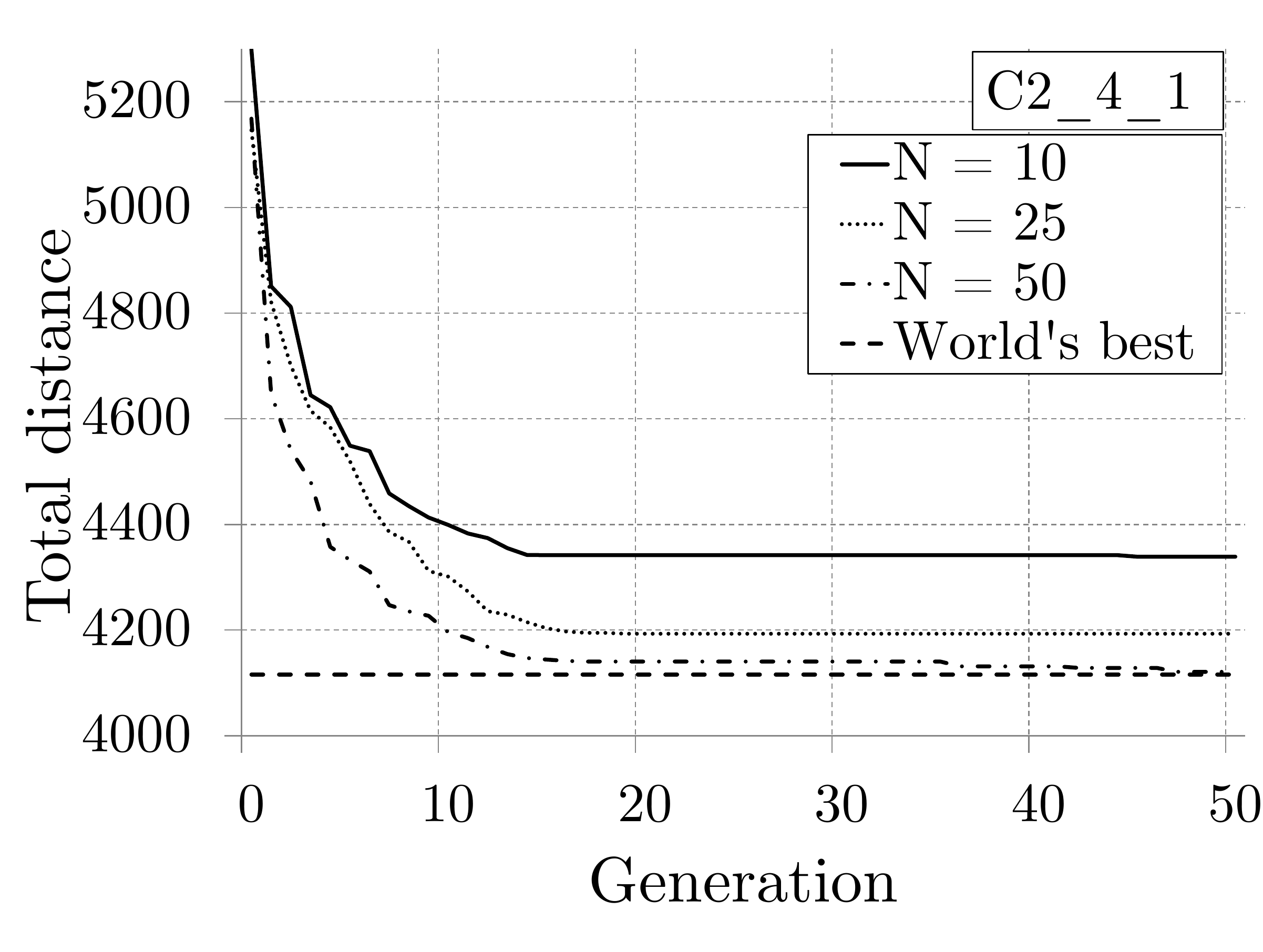}
\\

\end{tabular}

  \caption
  {
    Total distance traveled for various population sizes $N$ for tests (a) R1\_2\_2 and (b) C2\_4\_1.
  }
  \label{fig:various_N}
\end{figure}

The size of the population $N$ influences the execution time necessary to create a new generation of solutions. However, the probability of ending up with a set of similar individuals is lower in case of large populations. The problem of saturating the population is illustrated in Fig.~\ref{fig:various_N}. The experiments with the clustered customers have shown that the saturation of the population with similar individuals may occur relatively fast. The larger population should imply a larger population diversity. If the number of individuals with similar configurations exceeds a certain threshold, then the population is in the diversity crisis. Similarly, the number of children $N_{ch}$ generated for each pair of parents influences the quality of final solutions along with the execution time of the algorithm. Clearly, generating a larger number of children increases the probability of getting a well-fitted individual. On the other hand, the execution time of the algorithm increases with larger values of $N_{ch}$.

\begin{figure}[t!]
\centering

\renewcommand{\tabcolsep}{0cm}
\newcommand{\myfigwidth}{0.39}
\newcommand{\raiseshift}{0.1mm}
\newcommand{\mytextsize}{\scriptsize}

\begin{tabular}{ll}

a) & b) \\
\includegraphics[height=\myfigwidth\columnwidth]{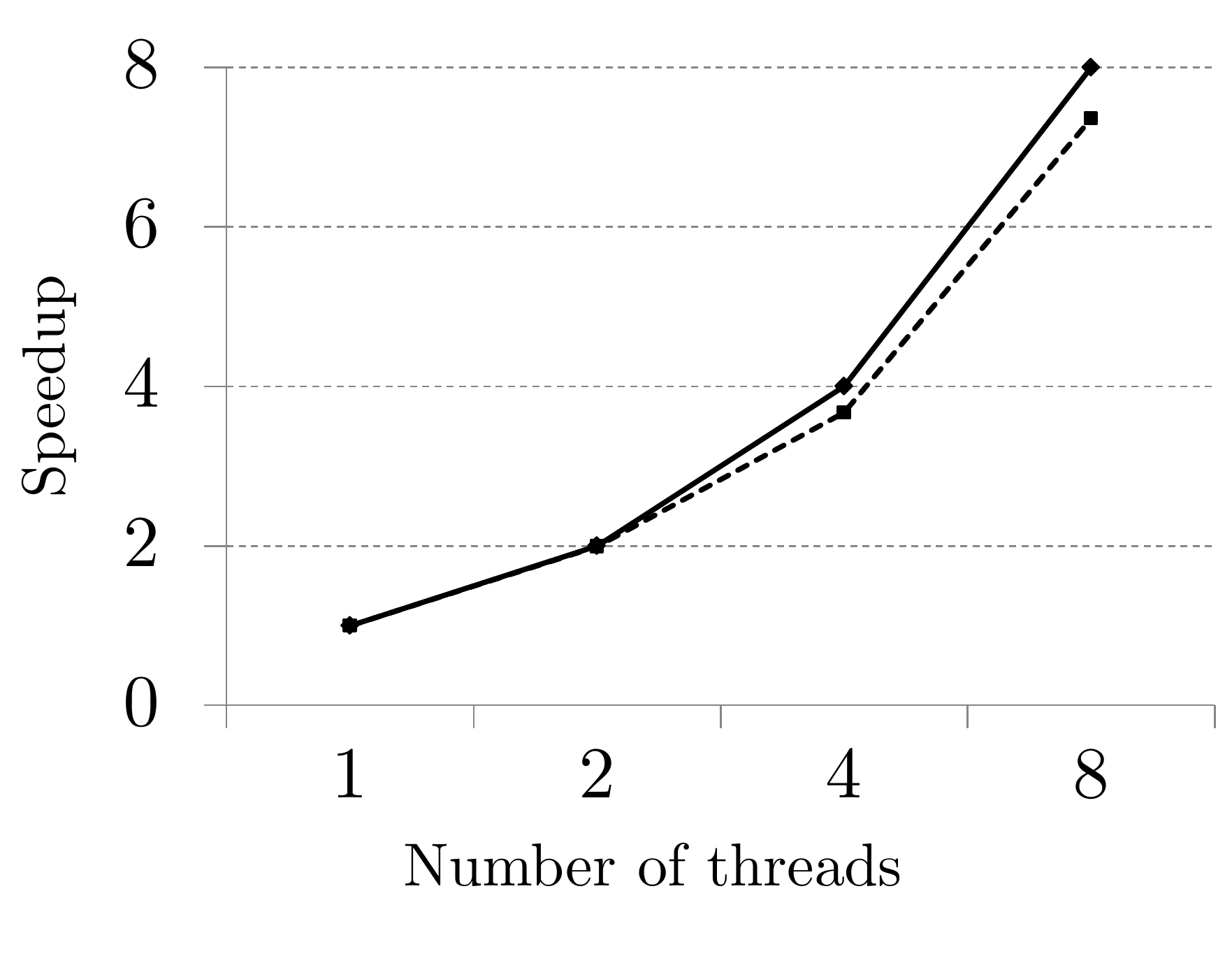} \hspace{0.01\columnwidth}&
\includegraphics[height=\myfigwidth\columnwidth]{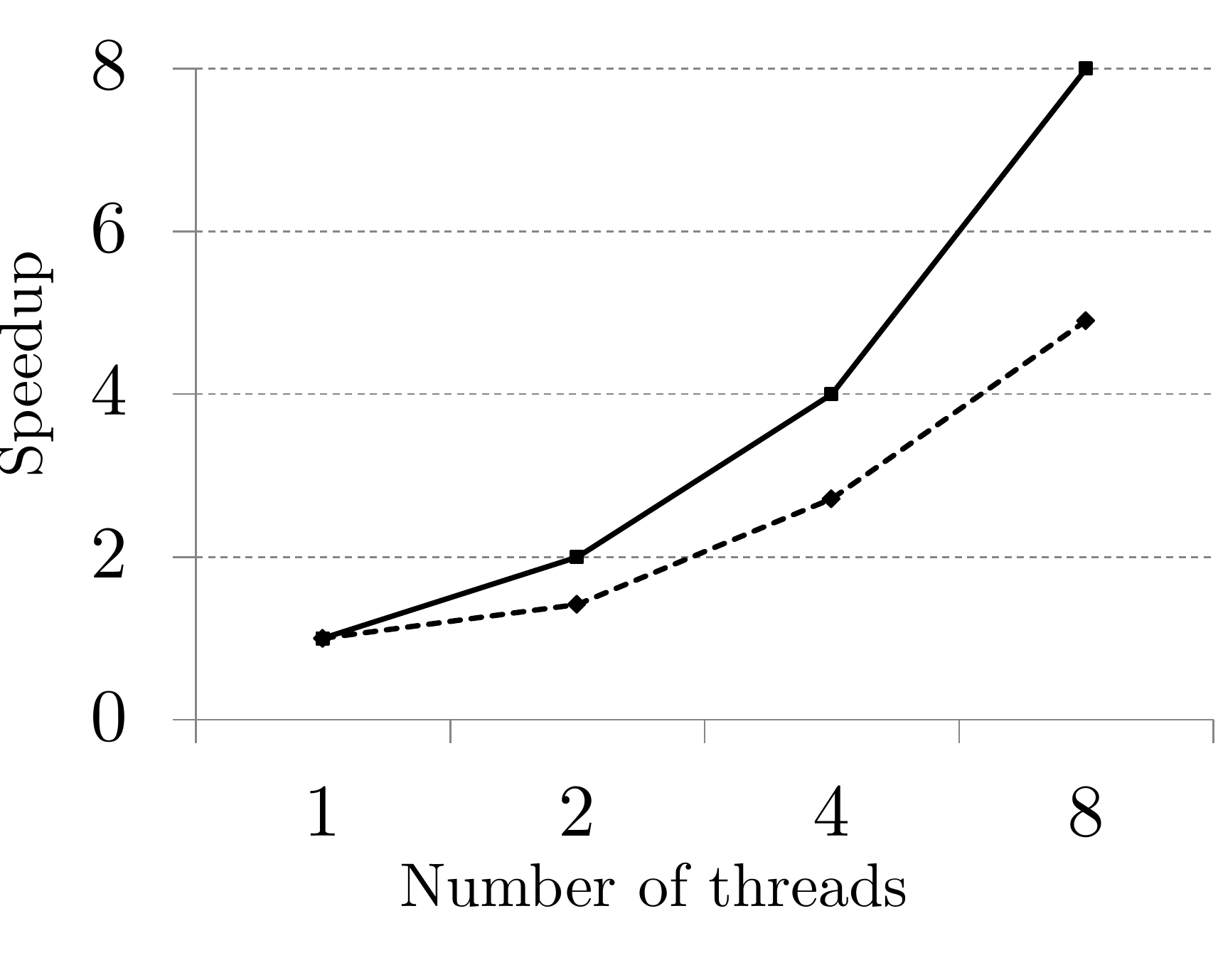}
\\

\end{tabular}

  \caption
  {
    Speedup $\mathcal{S}$ vs. number of threads $p$ for tests (a) RC1\_6\_3 and (b) C1\_2\_1 (continuous line shows the ideal speedup).
  }
  \label{fig:speed_up}
\end{figure}

The relative speedup obtained for two given problem instances is presented in Fig.~\ref{fig:speed_up}. The population size $N$ is usually larger than the number of threads $p$. The speedup depends not only on the problem size but also on its internal structure. If the number of generations required to obtain a minimal travel distance is large, then the relative speedup is almost ideal. However, if the solution converges to the minimum relatively fast, then the further improvements become difficult and the parallel overhead becomes more significant once the steady state is reached.

\section{Conclusions and Future Work} \label{sec:concl}

The parallel heuristic algorithm for minimizing the fleet size has proven to be effective and competitive by solving $90\%$ of problem instances to the current known optimum. The memetic algorithm for the distance minimization turned out to be powerful. A large number of parameters, both for the exploration and the exploitation of the search space, allow for adjusting the algorithm to the instance characteristics. The optimal assignment of parameters is to be cleared up during the further research. The experiments performed for various problem instances showed that the relative speedup is linear and close to the ideal one in many cases. The parallel algorithm significantly improved the world's best solution of the clustered Gehring and Homberger's test C1\_8\_2 containing 800 customers.

A two-stage approach of solving the VRPTW makes it possible to combine the presented algorithms with other well-known heuristics, e.g., simulated annealing or tabu search. We aim at determining the most effective and scalable combination of heuristics addressing both objectives of the VRPTW. Also, we plan to develop more efficient co-operation schemes for both stages of the parallel algorithm.

\subsubsection*{Acknowledgments.} We thank the following computing centers where
the computations of our project were carried out:
Academic Computer Centre in Gda\'{n}sk TASK,
Academic Computer Centre CYFRONET AGH,
Krak\'{o}w, Interdisciplinary Centre for
Mathematical and Computational Modeling,
Warsaw University, Wroc\l{}aw Centre for Networking
and Supercomputing.

\bibliographystyle{splncs}% bib style
\bibliography{ref_all_jn}% your bib database

\begin{thebibliography}{10}

\bibitem{Bard2002}
Bard, J.F., Kontoravdis, G., Yu, G.:
\newblock A branch-and-cut procedure for the vehicle routing problem with time
  windows.
\newblock Trans. Sc. \textbf{36}(2) (2002)  250--269

\bibitem{Irnich2006}
Irnich, S., Villeneuve, D.:
\newblock The shortest-path problem with resource constraints and k-cycle
  elimination.
\newblock INFORMS J. on Computing \textbf{18}(3) (2006)  391--406

\bibitem{Jepsen2006}
Jepsen, M., Petersen, B., Spoorendonk, S., Pisinger, D.:
\newblock A non-robust branch-and-cut-and-price algorithm for the vehicle
  routing problem with time windows.
\newblock Technical Report 06-03, University of Copenhagen, Denmark (2006)

\bibitem{Kallehauge2006}
Kallehauge, B., Larsen, J., Madsen, O.B.G.:
\newblock Lagrangian duality applied to the vehicle routing problem with time
  windows.
\newblock Comput. Oper. Res. \textbf{33}(5) (2006)  1464--1487

\bibitem{Chabrier2006}
Chabrier, A.:
\newblock Vehicle routing problem with elementary shortest path based column
  generation.
\newblock Comput. Oper. Res. \textbf{33}(10) (2006)  2972--2990

\bibitem{HombergerBenchmark}
{S}{I}{N}{T}{E}{F}:
\newblock Problems and benchmarks, {V}{R}{P}{T}{W}.
\newblock Website (2011)
  \url{http://www.sintef.no/Projectweb/TOP/VRPTW/Homberger-benchmark/}.

\bibitem{Nagata2010}
Nagata, Y., Br\"{a}ysy, O., Dullaert, W.:
\newblock A penalty-based edge assembly memetic algorithm for the vehicle
  routing problem with time windows.
\newblock Comput. Oper. Res. \textbf{37}(4) (2010)  724--737

\bibitem{Kallehauge2008Review}
Kallehauge, B.:
\newblock Formulations and exact algorithms for the vehicle routing problem
  with time windows.
\newblock Comput. Oper. Res. \textbf{35}(7) (2008)  2307--2330

\bibitem{Solomon1987}
Solomon, M.M.:
\newblock Algorithms for the vehicle routing and scheduling problems with time
  window constraints.
\newblock Oper. Res. \textbf{35} (1987)  254--265

\bibitem{Potvin1993}
Potvin, J.Y., Rousseau, J.M.:
\newblock A parallel route building algorithm for the vehicle routing and
  scheduling problem with time windows.
\newblock European J. of Oper. Res. \textbf{66}(3) (1993)  331--340

\bibitem{Pang2011}
Pang, K.W.:
\newblock An adaptive parallel route construction heuristic for the vehicle
  routing problem with time windows constraints.
\newblock Exp. Syst. Appl. \textbf{38}(9) (2011)  11939--11946

\bibitem{Thompson1993}
Thompson, P.M., Psaraftis, H.N.:
\newblock Cyclic transfer algorithms for multivehicle routing and scheduling
  problems.
\newblock Oper. Res. \textbf{41}(5) (1993)  935--946

\bibitem{Russell1995}
Russell, R.:
\newblock Hybrid heuristics for the vehicle routing problem with time windows.
\newblock Trans. Sc. \textbf{29}(2) (1995)  156

\bibitem{Potvin1995}
Potvin, J.Y., Rousseau, J.M.:
\newblock An exchange heuristic for routing problems with time windows.
\newblock J. of the Oper. Res. Society \textbf{46} (1995)  1433--1446

\bibitem{Zhong2007}
Zhong, Y., Pan, X.:
\newblock A hybrid optimization solution to {V}{R}{P}{T}{W} based on simulated
  annealing.
\newblock In: Automation and Logistics, 2007 {I}{E}{E}{E} International
  Conference on. (2007)  3113 --3117

\bibitem{Debudaj2005}
Debudaj-Grabysz, A., Czech, Z.J.:
\newblock A concurrent implementation of simulated annealing and its
  application to the {V}{R}{P}{T}{W} optimization problem.
\newblock In: Distributed and Parallel Systems. Volume 777 of The Kluwer
  International Series in Engineering and Computer Science.
\newblock Springer US (2005)  201--209

\bibitem{Li2013}
Li, Y.:
\newblock An improved simulated annealing algorithm and its application in the
  logistics distribution center location problem.
\newblock Applied Mechanics and Materials \textbf{389} (2013)  990--994

\bibitem{Cordeau2001}
Cordeau, J., Laporte, G., Hautes, E., Commerciales, E., Gerad, L.C.D.:
\newblock A unified tabu search heuristic for vehicle routing problems with
  time windows.
\newblock J. of the Oper. Res. Society \textbf{52} (2001)  928--936

\bibitem{Sin2002}
Ho, S.C., Haugland, D.:
\newblock A tabu search heuristic for the vehicle routing problem with time
  windows and split deliveries.
\newblock Comp. and Oper. Res. \textbf{31} (2002)  1947--1964

\bibitem{Xuan2006}
Tan, X., Zhuo, X., Zhang, J.:
\newblock Ant colony system for optimizing vehicle routing problem with time
  windows ({V}{R}{P}{T}{W}).
\newblock In: Proc. of the 2006 int. conf. on Comp. Intell. and Bioinf.
  ICIC'06, Berlin, Heidelberg, Springer-Verlag (2006)  33--38

\bibitem{Qi2008}
Qi, C., Sun, Y.:
\newblock An improved ant colony algorithm for {V}{R}{P}{T}{W}.
\newblock In: Proc. of the 2008 Int. Conf. on Comp. Sc. and Soft. Eng. (2008)
  455--458

\bibitem{Braysy2005Survey}
Br\"{a}ysy, O., Gendreau, M.:
\newblock {V}ehicle {R}outing {P}roblem with {T}ime {W}indows, part {I}{I}:
  Metaheuristics.
\newblock Trans. Sc. \textbf{39}(1) (2005)  119--139

\bibitem{Cheng2009}
Cheng, C.B., Wang, K.P.:
\newblock Solving a vehicle routing problem with time windows by a
  decomposition technique and a genetic algorithm.
\newblock Expert Syst. Appl. \textbf{36}(4) (2009)  7758--7763

\bibitem{Kamkar2010}
Kamkar, I., Poostchi, M., Akbarzadeh~Totonchi, M.R.:
\newblock A cellular genetic algorithm for solving the vehicle routing problem
  with time windows.
\newblock In: Soft Comp. in Industrial App. Volume~75 of Advances in
  Intelligent and Soft Computing.
\newblock Springer, Berlin, Heidelberg (2010)  263--270

\bibitem{Ursani2011}
Ursani, Z., Essam, D., Cornforth, D., Stocker, R.:
\newblock Localized genetic algorithm for vehicle routing problem with time
  windows.
\newblock Appl. Soft Comput. \textbf{11}(8) (2011)  5375--5390

\bibitem{Gehring2002}
Gehring, H., Homberger, J.:
\newblock Parallelization of a two-phase metaheuristic for routing problems
  with time windows.
\newblock Journal of Heuristics \textbf{8}(3) (2002)  251--276

\bibitem{Mester2005}
Mester, D., Br\"{a}ysy, O.:
\newblock Active guided evolution strategies for large-scale vehicle routing
  problems with time windows.
\newblock Comput. Oper. Res. \textbf{32}(6) (2005)  1593--1614

\bibitem{Kanoh2010}
Kanoh, H., Tsukahara, S.:
\newblock Solving real-world vehicle routing problems with time windows using
  virus evolution strategy.
\newblock Int. J. Know.-Based Intell. Eng. Syst. \textbf{14}(3) (2010)
  115--126

\bibitem{Berger2004}
Berger, J., Barkaoui, M.:
\newblock A parallel hybrid genetic algorithm for the vehicle routing problem
  with time windows.
\newblock Comp. and Oper. Res. (2004)  2037--2053

\bibitem{Labadi2008}
Labadi, N., Prins, C., Reghioui, M.:
\newblock A memetic algorithm for the vehicle routing problem with time
  windows.
\newblock Oper. Res. \textbf{42}(3) (2008)  415--431

\bibitem{Nagata2009}
Nagata, Y., Br{\"a}ysy, O.:
\newblock A powerful route minimization heuristic for the vehicle routing
  problem with time windows.
\newblock Oper. Res. Lett. \textbf{37}(5) (2009)  333--338

\bibitem{Blocho2010}
Blocho, M., Czech, Z.J.:
\newblock An improved route minimization algorithm for the vehicle routing
  problem with time windows.
\newblock Studia {I}nformatica \textbf{32}(99) (2010)  5--19

\bibitem{Nalepa2012TAAI}
Nalepa, J., Czech, Z.J.:
\newblock Adaptive threads co-operation schemes in a parallel heuristic
  algorithm for the vehicle routing problem with time windows.
\newblock Theoretical and {A}pplied {I}nformatics \textbf{24}(3) (2012)
  191--203

\bibitem{Kawulok2012SPR}
Kawulok, M., Nalepa, J.:
\newblock Support vector machines training data selection using a genetic
  algorithm.
\newblock In Gimel’farb, G., Hancock, E., Imiya, A., Kuijper, A., Kudo, M.,
  Omachi, S., Windeatt, T., Yamada, K., eds.: Structural, Syntactic, and
  Statistical Pattern Recognition. Volume 7626 of Lecture Notes in Computer
  Science.
\newblock Springer Berlin Heidelberg (2012)  557--565

\bibitem{galera}
:
\newblock {CI} {T}{A}{S}{K}.
\newblock Website (2011) \url{http://www.task.gda.pl/kdm/sprzet/Galera}.

\end{thebibliography}

\end{document}